\documentclass[american,english,jcp, twocolumn]{revtex4-1}
\usepackage[T1]{fontenc}
\usepackage[latin9]{inputenc}
\usepackage{bm}
\usepackage{amsmath}
\usepackage{amssymb}
\usepackage{graphicx}

\makeatletter

\@ifundefined{textcolor}{}
{%
 \definecolor{BLACK}{gray}{0}
 \definecolor{WHITE}{gray}{1}
 \definecolor{RED}{rgb}{1,0,0}
 \definecolor{GREEN}{rgb}{0,1,0}
 \definecolor{BLUE}{rgb}{0,0,1}
 \definecolor{CYAN}{cmyk}{1,0,0,0}
 \definecolor{MAGENTA}{cmyk}{0,1,0,0}
 \definecolor{YELLOW}{cmyk}{0,0,1,0}
}

\usepackage{amsmath}
\usepackage{subfigure}

\makeatother

\usepackage{babel}
\begin{document}

\title{Brownian microhydrodynamics of active filaments}

\author{Abhrajit Laskar}

\email{abhra@imsc.res.in}

\selectlanguage{english}%

\affiliation{The Institute of Mathematical Sciences, CIT Campus, Chennai 600113, India}

\author{R. Adhikari}

\email{rjoy@imsc.res.in}

\selectlanguage{english}%

\affiliation{The Institute of Mathematical Sciences, CIT Campus, Chennai 600113, India}
\begin{abstract}
Slender bodies capable of spontaneous motion in the absence of external actuation
in an otherwise quiescent fluid are common in biological, physical and technological
contexts. The interplay between the spontaneous fluid flow, Brownian motion, and
the elasticity of the body presents a challenging fluid-structure interaction problem.
Here, we model this problem by approximating the slender body as an elastic filament
that can impose non-equilibrium velocities or stresses at the fluid-structure interface.
We derive equations of motion for such an active filament by enforcing momentum conservation
in the fluid-structure interaction and assuming slow viscous flow in the fluid. The
fluid-structure interaction is obtained, to any desired degree of accuracy, through
the solution of an integral equation. A simplified form of the equations of motion,
that allows for efficient numerical solutions, is obtained by applying the Kirkwood-Riseman
superposition approximation to the integral equation. We use this form of the equation
of motion to study dynamical steady states in free and hinged minimally active filaments.
Our model provides the foundation to study collective phenomena in momentum-conserving,
Brownian, active filament suspensions.
\end{abstract}
\maketitle

\section{introduction}

Slender bodies capable of spontaneous motion in viscous fluids are common in biological,
chemical, physical and technological contexts. Examples from biology, in increasing
degree of molecular complexity, include microtubules driven by molecular motors,
axonemes, cilia, flagella \cite{sanchez2011,sanchez2012,berg2000,ishiwata1991,lindemann1972,machin1958,camalet1999,camalet2000,bessen1980,nedelec1997,camazine2003,kruse2004},
their synchronization and metachronal wave \cite{brumley2014flagellar,brumley2012hydrodynamic}.
In chemistry and physics, self-assembled bundles of microtubules driven by kinesin
motors yields a model experimental system in which broken symmetry, collective excitations,
and topological defects can be studied out of equilibrium \cite{sanchez2011,sanchez2012}.
In technology, much recent research has been directed towards the synthesis of slender
bodies capable of spontaneous motion \cite{wang2013,paxton2004,vicario2005}. Such
self-actuated slender bodies are expected to have many microfluidic and biomimetic
applications \cite{williams2014self}.

Despite the great variety in both the structure of the body and its mechanism of
self-actuation, the examples above have three features in common: the spontaneous
motion of the slender body produces flow in the ambient fluid, the body is of a size
sufficiently small to make Brownian fluctuations important, and the body resists
deformation produced by the spontaneous flow and Brownian fluctuations. Any universal
emergent behaviour in active slender bodies must appear from the interplay between
fluid flow, Brownian fluctuations and the elasticity. Such systems present a new
class of fluid-structure interaction problems.

In this paper, we construct a theory of active slender bodies, by modeling them as
filaments that enforce slip velocities or non-equilibrium stresses at the fluid-structure
boundary. A multitude of microscopic mechanisms can produce such velocities or stresses.
Our theory isolates the specific microscopic details of self-actuation in the boundary
conditions, from which universal, macroscopic fluid flow can be generated. Such flow
results from the exchange of momentum between the body and the fluid, and since no
external forces act on the body or the fluid, the sum of their linear momenta is
conserved. In the absence of external torques, the sum of their angular momenta is
similarly conserved. These two constraints are explicitly taken into account when
computing the fluid flow within our theory. 

The flow is computed through a discretization which replaces the continuous filament
by a chain of spheres connected by non-linear springs. The spheres produce spontaneous
hydrodynamic flow while the springs penalize changes in filament length and filament
curvature. The antecedent of such a bead-spring discretization of a continuous filament
traces back to Kramers, who used it to model the dynamics of a polymer. The crucial
difference between the model of Kramers and our adaptation of it is that the spheres
in our theory produce spontaneous flows. Each sphere must, therefore, be active.
Such chains of active spheres have been used previously to model active filaments.
In the earliest such model \cite{jayaraman2012,laskar2013}, the spheres produce
dipolar stresslet flows but are individually non-motile and are assumed so large
that Brownian effects are negligible. In a subsequent contribution \cite{chelakkot2014flagellar},
the spheres are taken to be motile, Brownian effects are included in two-dimensions,
but contributions from non-local hydrodynamic flow are neglected. In a related model,
passive spheres are driven by tangential active stresses, hydrodynamically correlated
Brownian motion is included in three-dimensions, but active flow is neglected \cite{jiang2014hydrodynamic,jiang2014motion}.
Our theory presented here contains all previous models as special cases. 

In our recent work \cite{singh2014many}, the problem of computing the fluid flow
of $N$ active bodies has been solved by transforming the local conservation law
for momentum, which under the conditions relevant to the microhydrodynamic regime
is the Stokes equation, into an integral equation over the sphere boundaries. The
solution of this boundary integral equation gives the rigid body motion of the active
spheres, constrained by the conservation of linear and angular momentum, as a linear
function of the forces, torques and the active boundary condition: \begin{subequations}\label{eq: singh_propulsionmatrix}
\begin{alignat}{1}
\mathbf{V}_{n} & =\sum_{m=1}^{N}\left[\bm{\mu}_{nm}^{TT}\cdot\mathbf{F}_{m}^{e}+\bm{\mu}_{nm}^{TR}\cdot\mathbf{T}_{m}^{e}\right]\nonumber \\
 & +\sum_{m\ne n}^{N}\sum_{l=1}^{\infty}\left[\bm{\pi}_{nm}^{(T,\, l+1)}\odot\mathbf{V}_{m}^{(l+1)}\right]+\mathbf{V}_{n}^{a},\\
\bm{\Omega}_{n} & =\sum_{m=1}^{N}\left[\bm{\mu}_{nm}^{RT}\cdot\mathbf{F}_{m}^{e}+\bm{\mu}_{nm}^{RR}\cdot\mathbf{T}_{m}^{e}\right]\nonumber \\
 & +\sum_{m\ne n}^{N}\sum_{l=1}^{\infty}\left[\bm{\pi}_{nm}^{(R,\, l+1)}\odot\mathbf{V}_{m}^{(l+1)}\right]+\bm{\Omega}_{n}^{a}.
\end{alignat}
\end{subequations} In the above, $\mathbf{V}_{n}$ and $\bm{\Omega}_{n}$ are the
velocity and angular velocity of the $n$-th particle, $\mathbf{V}_{n}^{a}$ and
$\bm{\Omega}_{n}^{a}$ are the self-propulsion and self-rotation contributions to
the rigid body motion, the $\bm{\mu}$ are the usual mobility matrices that relate
external forces and torques to the rigid body motion and the $\bm{\pi}$ are propulsion
matrices \cite{singh2014many} that relate $\mathbf{V}_{m}^{(l+1)}$, the $l$-th
mode of the non-equilibrium slip velocity on the surface of the $m$-th particle,
to the rigid body motion (see below). These relations clearly show that rigid body
motion of active particles is possible in the absence of external forces and torques,
$\mathbf{F}_{m}^{e}=0$, $\mathbf{T}_{m}^{e}=0$, and even in the absence of self-propulsion,
$\mathbf{V}_{n}^{a}=0$ and self-rotation $\bm{\Omega}_{n}^{a}=0$. Propulsion matrices,
and not mobility matrices, are the key quantities that describe the correlated motion
of active particles in a viscous fluid, constrained by the conservation of momentum
and angular momentum \cite{singh2014many}. While mobility and propulsion matrices
have to be computed numerically for particles of general shape, analytical expressions
can be derived when the particles are spheres \cite{singh2014many}. 

Here, we use mobility and propulsion matrices for spheres of radius $a$, computed
in the superposition approximation first introduced by Kirkwood and Riseman \cite{kirkwood1948intrinsic},
again, in the context of the dynamics of a polymer. In this approximation, the mobility
matrices reduce to the well-known Rotne-Prager-Yamakawa tensors while the propulsion
matrices are obtained analytically \cite{singh2014many} as gradients of the fundamental
solution of the Stokes equation for an unbounded fluid. For a passive polymer, Yoshizaki
and Yamakawa \cite{Yoshizaki1980} verified that the superposition approximation
is correct to $\mathcal{O}((a/b)^{3})$, where $b$ is the mean separation between
spheres, as $N\rightarrow\infty$ . Since the propulsion matrices decay more rapidly
with separation than mobility matrices, the superposition approximation for active
filaments is also accurate to $\mathcal{O}((a/b)^{3})$ \cite{singh2014many}. The
resulting equations of motion are used to study the dynamics of active filaments
where the spheres produce dipolar flows and, so, are not individually motile. This
part of our work complements studies of active filaments consisting of individually
motile particles. We emphasis that our general theory includes both the motile and
non-motile cases studied previously.

In the next section we present a generalization of the theory of Brownian motion
of hydrodynamically interacting spherical particles to include surface activity.
In section III, we construct equations of motion for active filaments using the results
of this general theory. In section IV we introduce a minimal model for an active
filament by discarding all but leading terms in the filament equations of motion
derived previously. Here we also study dynamics of such active filaments when both
ends of the filament are free and when one end is tethered and the other end is free.
In the first case, for sufficient strength of activity the filament is unstable to
transverse perturbations, which results in a sequence of translational, rotational
and oscillatory dynamical steady states. In the second case, we find a sequence of
rotational and oscillatory dynamical steady states, both of which are limit cycles
in the phase space of the dynamical system described by the equations of motion.
In section V, linear stability analysis shows that the transition to a dynamic state
happens due a simple instability in both cases. This analysis shows that non-local
hydrodynamic interactions are essential for the dynamic instability to occur. We
conclude with a discussion of the application of our equations of motion to study
collective phenomena in suspensions of active filaments \cite{pandey2014flow,saintillan2013}
and other soft dissipative structures.

\section{Brownian Microhydrodynamics of Active Spheres}

We consider the motion of $N$ spherical active particles of radius $a$ in an incompressible
viscous fluid. The center of the $n$-th sphere is at $\mathbf{R}_{n}$ and its orientation
is specified by the unit vector $\mathbf{p}_{n}$. The fluid can exchange both momentum
and angular momentum with the particles, of amounts determined by integrals of the
momentum flux over the particle boundaries. In addition to this contact contribution,
the particles may be acted on by body forces and torques. At low Reynolds numbers
inertia, of both the particles and the fluid, can be neglected, which results in
an instantaneous balance of surface and body contributions to forces and torques.
Newton's laws, therefore, reduce to a pair of constraints on the fluid stress at
the surface of every particle, \begin{subequations}
\begin{alignat}{1}
M{\bf \dot{V}}_{n} & =\oint\mathbf{n}\cdot\bm{\sigma}dS_{n}+\mathbf{F}_{n}^{e}=0,\\
I\bm{\dot{\Omega}}_{n} & =\oint\mathbf{r\times n}\cdot\bm{\sigma}dS_{n}+\mathbf{T}_{n}^{e}=0.
\end{alignat}
\end{subequations}The fluid stress $\bm{\sigma}=-p\mathbb{I}+\eta(\bm{\nabla}\mathbf{v}+\bm{\nabla}\mathbf{v}^{T})$
has both hydrostatic and viscous contributions and is determined from conditions
of incompressibility and local momentum conservation
\begin{gather}
\bm{\nabla}\cdot{\bf v}=0;\quad\bm{\nabla}\cdot\bm{\sigma}=0.
\end{gather}
where $\mathbf{v}$ is the fluid velocity, $p$ is pressure and $\eta$ is the viscosity.
The solution of this Stokes system provides the stress, from which the contact contribution
of the force and torque on the every particle can be determined. In the absence of
inertia and body forces, the conservation of particle momentum requires that the
net contact force and the net contact torque on every particle be zero. The solution
of the Stokes equation is determined by the boundary conditions on the particle surfaces
and at infinity. While activity can be expressed through a variety of boundary conditions
on both the fluid velocity and on the fluid stress , we assume here an active slip
at the surface of the particle \cite{singh2014many}. This encompasses a wide variety
of active phenomena, including electrophoresis, diffusiophoresis \cite{anderson1991},
self-phoresis due to chemical catalysis \cite{Snigdha2014RingClosure}, and even
swimming of microorganisms \cite{blake1971a,ishikawa2006}. We chose the fluid to
be at rest at infinity. The boundary conditions, therefore, are\begin{subequations}
\begin{gather}
\mathbf{v}(\mathbf{R}_{n}+\bm{\rho}_{n})=\mathbf{V}_{n}+\bm{\Omega}_{n}\times\bm{\rho}_{n}+\mathbf{v}^{a}(\bm{\rho}_{n})\\
\,|\mathbf{v}|\rightarrow0,\quad|p|\rightarrow0,\quad|\mathbf{r}|\rightarrow\infty.
\end{gather}
\end{subequations}

The task of obtaining the solution of the Stokes equation is substantially simplified
by recognizing that the three-dimensional partial differential equation can be reduced,
instead, to a two-dimensional integral equation over the particle boundaries. The
starting point of this development is the integral representation of Stokes flow
\cite{ladyzhenskaya1969,pozrikidis1992},
\begin{alignat}{1}
8\pi\eta v_{i}(\mathbf{r})= & -\sum_{m=1}^{N}\int G_{ij}(\mathbf{r},\mathbf{r}_{m})f_{j}(\mathbf{r}_{m})dS_{m}\nonumber \\
+ & \ \eta\sum_{m=1}^{N}\int K_{jik}(\mathbf{r},\mathbf{r}_{m})n_{k}v_{j}(\mathbf{r}_{m})dS_{m}
\end{alignat}
which provides the solution of Stokes equation in terms of the velocity and the traction,
$\mathbf{f}=\mathbf{n}\cdot\bm{\sigma}$, over the particle boundaries. The kernels
in the integral representation are the Green's function $\mathbf{G}$, the pressure
vector $\mathbf{p}$ and the stress tensor, $\mathbf{K}$. \begin{subequations}
\begin{alignat}{1}
p_{i}(\mathbf{r},\mathbf{r'}) & =-\nabla_{i}\bm{\bm{\nabla}}^{2}\rho=\frac{\rho_{i}}{\rho^{3}}\\
G_{ij}(\mathbf{r},\mathbf{r'}) & =\left(\bm{\bm{\nabla}}^{2}\delta_{ij}-\nabla_{i}\nabla_{j}\right)\rho=\frac{\delta_{ij}}{\rho}+\frac{\rho_{i}\rho_{j}}{\rho^{3}};\\
K_{ijk}(\mathbf{r},\mathbf{r'}) & =-\delta_{ik}p_{j}+\nabla_{k}G_{ij}+\nabla_{i}G_{jk}.
\end{alignat}
\end{subequations}

In the absence of boundaries, these kernels are translationally invariant and, so,
are functions of the separation $\boldsymbol{\rho}=\mathbf{r}-\mathbf{r'}$. The
boundary integrals can be evaluated analytically by expanding the boundary fields
in complete orthogonal bases \cite{singh2014many,ghose2014a}, which are most conveniently
chosen to be the tensorial spherical harmonics, $\mathbf{Y}^{(l)}$
\begin{gather}
\mathbf{Y}_{\alpha_{1}\alpha_{2}\dots\alpha_{l}}^{(l)}(\bm{\hat{\rho}})=(-1)^{l+1}\rho^{l+1}\bm{\nabla}_{\alpha_{1}}\dots\bm{\nabla}_{\alpha_{l}}\left(\frac{1}{\rho}\right).
\end{gather}
In this basis, the surface velocity and traction on each particle is expanded as
\cite{ladd1988,ghose2014a}\begin{subequations}
\begin{alignat}{1}
\mathbf{f}\,(\mathbf{R}_{m}+\bm{\rho}_{m}) & =\sum_{l=0}^{\infty}\frac{2l+1}{4\pi a^{2}}\mathbf{F}_{m}^{(l+1)}\odot\mathbf{Y}^{(l)}(\bm{\hat{\rho}}_{m}),\\
\mathbf{v}\,(\mathbf{R}_{m}+\bm{\rho}_{m}) & =\sum_{l=0}^{\infty}\frac{1}{l!(2l-1)!!}\mathbf{V}_{m}^{(l+1)}\odot\mathbf{Y}^{(l)}(\bm{\hat{\rho}}_{m}),
\end{alignat}
\end{subequations} where the orthogonality of the basis functions yields the expansion
coefficients as \begin{subequations}
\begin{alignat}{1}
\mathbf{F}_{m}^{(l+1)} & =\frac{1}{l!(2l-1)!!}\int\mathbf{f}(\mathbf{R}_{m}+\bm{\rho}_{m})\mathbf{Y}^{(l)}(\bm{\hat{\rho}}_{m})dS_{m},\\
\mathbf{V}_{m}^{(l+1)} & =\frac{2l+1}{4\pi a^{2}}\int\mathbf{v}(\mathbf{R}_{m}+\bm{\rho}_{m})\mathbf{Y}^{(l)}(\bm{\hat{\rho}}_{m})dS_{m}.
\end{alignat}
\end{subequations}These expansion coefficients are tensors of rank $l+1$, symmetric
and irreducible in their last $l$ indices \cite{singh2014many,ghose2014a}. The
forces, torques, velocities and angular velocities are obtained from
\begin{alignat*}{1}
\mathbf{F}_{m}^{(1)} & =-\mathbf{F}_{m}^{e};\:2\epsilon\cdot a\mathbf{F}_{m}^{(2)}=\mathbf{T}_{m}^{e}\\
\mathbf{V}_{m}^{(1)} & =\mathbf{V}_{m}-\langle\mathbf{v}_{m}^{a}\rangle;\:2\epsilon\cdot\mathbf{V}_{m}^{(2)}=-a\bm{\Omega}_{m}+\dfrac{3}{2a}\langle\bm{\rho}_{m}\times\mathbf{v}_{m}^{a}\rangle
\end{alignat*}
where the angle brackets indicate integration of the enclosed term over the surface
of the sphere and dividing the result by the surface area. Inserting the expansions
in the boundary integral representation leads to a succinct expression for the fluid
flow in terms of the expansion coefficients \cite{singh2014many},
\begin{alignat}{1}
8\pi\eta\mathbf{v}(\mathbf{r})= & -\sum_{m=1}^{N}\sum_{l=0}^{\infty}\mathbf{G}^{(l+1)}(\mathbf{r},\mathbf{R}_{m})\odot\mathbf{F}_{m}^{(l+1)}\nonumber \\
+ & \;\eta\sum_{m=1}^{N}\sum_{l=1}^{\infty}\mathbf{K}^{(l+1)}(\mathbf{r},\mathbf{R}_{m})\odot\mathbf{V}_{m}^{(l+1)}.
\end{alignat}
The boundary integrals $\mathbf{G}^{(l)}$ and $\mathbf{K}^{(l)}$ can be evaluated
explicitly in terms of the Green's function $\mathbf{G}$ and its derivatives. In
this expression, the velocity coefficients can be computed from the boundary condition,
but the traction coefficients remain unknown. To determine the traction coefficients,
the boundary condition is first enforced on the boundary of $n$-th particle, the
resulting equation is weighted by the $l$-th tensorial harmonic and finally integrated
over the $n$-th boundary. This Galerkin procedure yields an infinite-dimensional
linear system of equations for the unknown traction coefficients \cite{singh2014many},
\begin{alignat}{1}
4\pi\eta\mathbf{V}_{n}^{(l+1)} & =-\sum_{m=1}^{N}\sum_{l=0}^{\infty}\mathbf{G}_{nm}^{(l+1,l'+1)}(\mathbf{R}_{n},\mathbf{R}_{m})\odot\mathbf{F}_{m}^{(l'+1)}\nonumber \\
 & +\;\eta\sum_{m=1}^{N}\sum_{l=1}^{\infty}\mathbf{K}_{nm}^{(l+1,l'+1)}(\mathbf{R}_{n},\mathbf{R}_{m})\odot\mathbf{V}_{m}^{(l'+1)}.\label{eq:RBM_matrix}
\end{alignat}
where the matrix elements $\mathbf{G}_{nm}^{(l,l')}$ and $\mathbf{K}_{nm}^{(l,l')}$
can, again, be evaluated analytically in terms of the Green's function $\mathbf{G}$
and its derivatives \cite{singh2014many}. The formal solution of this linear system
yields Eq. \ref{eq: singh_propulsionmatrix}, which expresses the rigid body motion
in terms of the body forces and body torques and the coefficients of the active slip
velocity. These equations reflect the linearity of both the Stokes equations and
the boundary conditions. The hydrodynamically correlated motion is a sum of passive
terms, proportional to the external body forces and torques, and active terms, proportional
the modes of the active velocity. The last terms in each of the equations are the
self-propulsion and self-rotation velocities of an isolated active sphere, which
are given in terms of the active slip velocity as \begin{subequations}
\begin{alignat}{1}
4\pi a^{2}\mathbf{V}_{n}^{a} & =-\int\mathbf{v}^{a}(\bm{\rho}_{n})dS_{n}\\
4\pi a^{2}\bm{\Omega}_{n}^{a} & =-\frac{3}{2a^{2}}\int\bm{\rho}_{n}\times\mathbf{v}^{a}(\bm{\rho}_{n})dS_{n}
\end{alignat}
\end{subequations}The above two equations have been known in the context of phoresis
\cite{anderson1989} and swimming by surface distortions \cite{felderhof2015stokesian}.
They were later derived by the use of the Lorentz reciprocal relation \cite{stone1996}.
The work of \cite{ghose2014a} derives these from the boundary integral representation
of Stokes flow. The propulsion matrices emerge naturally from the solution of the
boundary integral equation as coefficients that determine the active contribution
to the hydrodynamic interaction. Thus, Eq. \ref{eq: singh_propulsionmatrix} expresses
hydrodynamic interactions between active particles acted upon by external forces
and torques \cite{singh2014many}. 

We note that the boundary integral method yields the hydrodynamic interaction between
particles without the need to resolve bulk fluid degrees of freedom. This makes Eq.
1 especially useful for computing the hydrodynamic interaction of active particles
in three dimensions, as the computational cost of resolving fluid degrees of freedom
is completely eliminated \cite{singh2014many,singh2014pystokes}. \foreignlanguage{american}{}
\begin{figure*}
\includegraphics[width=1\textwidth]{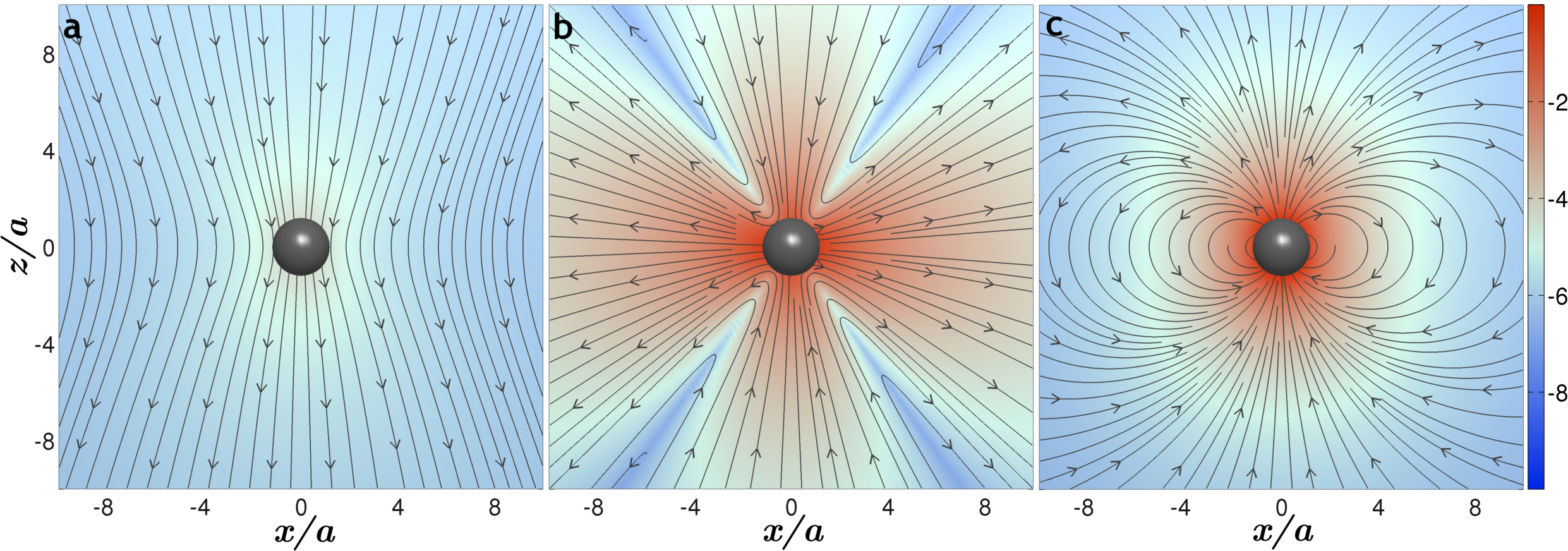}

\protect\caption{Streamlines of irreducible modes of slow viscous flow around an active sphere. Displayed
in (a) is $\mathcal{F}^{0}\mathbf{G}\cdot\mathbf{F}$, the flow around a sphere translating
under the action of a force $\mathbf{F}$; in (b) is $\mathcal{F}^{1}\bm{\nabla}\mathbf{G}\odot\mathbf{V}^{(2)}$,
the minimal flow around a non-motile active sphere; and in (c) is $\bm{\nabla}^{2}\mathbf{G}\odot\mathbf{V}^{(3)}$,
the minimal flow around a motile active sphere. In the limit $a\rightarrow0,$ the
flows in (a) and (b) reduce to a Stokeslet and stresslet, respectively, while (c)
is a degenerate velocity quadrupole. The background colour is proportional to the
logarithm of the velocity magnitude. }
\end{figure*}

The addition of Brownian fluctuations to Eq. \ref{eq: singh_propulsionmatrix} is
accomplished by appeal to linearity, the balance of dissipation and fluctuation and
Onsager symmetry of the mobilities. The generalization of the Einstein-Smoluchoswki
description of the diffusion of a passive colloidal particle to the hydrodynamically
correlated diffusive motion of $N$ colloidal particles was completed by several
authors using Liouville, kinetic theory, Fokker-Planck and Langevin approaches \cite{Murphy1972}.
The Langevin approach provides the most direct way of incorporating in Eq. \ref{eq: singh_propulsionmatrix}
by promoting them to a set of stochastic differential equations with a state-dependent
fluctuation. The fluctuations are chosen to compensate each source of dissipation
such that distribution of positions and orientations is Gibbsian in the absence of
activity. The fluctuations, then, are correlated Wiener processes with variances
proportional to the mobility matrices. The positivity of dissipation ensures that
mobility matrices are positive-definite and Onsager symmetry constrains them to be
symmetric in both the particle and translation-rotation indices. These two properties
ensure that a mobility matrix can be factorised into a lower triangular matrix and
its transpose, any one of which is a ``square-root'' of the mobility matrix. The
fluctuations can then be expressed as products of uncorrelated Wiener processes $\bm{\xi}^{T},\bm{\xi}^{R},\bm{\eta}^{T},\bm{\eta}^{R}$
and the ``square-root'' Cholesky factors. By linearity, dissipative, Brownian and
active motions are additive. Therefore, the generalization of Brownian hydrodynamics
to $N$ active particles, expressed as coupled Langevin equations, is

\begin{widetext}\begin{subequations}

\begin{alignat}{1}
\negthickspace\negmedspace\mathbf{V}_{n} & =\sum_{m=1}^{N}\underbrace{\bm{\mu}_{nm}^{TT}\cdot\mathbf{F}_{m}+\bm{\mu}_{nm}^{TR}\mathbf{\cdot T}_{m}}_{Passive}+\sum_{m=1}^{N}\underbrace{\sqrt{2k_{B}T\bm{\mu}_{nm}^{TT}}\cdot\ \bm{\xi}_{m}^{T}+\sqrt{2k_{B}T\bm{\mu}_{nm}^{TR}\cdot}\ \bm{\xi}_{m}^{R}}_{Brownian}+\sum_{m\neq n}^{N}\,\sum_{l=1}^{\infty}\underbrace{\bm{\pi}_{nm}^{(T,\, l+1)}\odot{\bf V}_{m}^{(l+1)}+\mathbf{V}_{n}^{a}}_{Active}\label{eq:flowInTermsofPropulsionMatrix}\\
\negthickspace\negmedspace\bm{\Omega}_{n} & =\sum_{m=1}^{N}\underbrace{\bm{\mu}_{nm}^{RT}\mathbf{\cdot F}_{m}+\bm{\mu}_{nm}^{RR}\cdot\mathbf{T}_{m}}_{Passive}+\sum_{m=1}^{N}\underbrace{\sqrt{2k_{B}T\bm{\mu}_{nm}^{RT}\cdot}\ \bm{\eta}_{m}^{T}+\sqrt{2k_{B}T\bm{\mu}_{nm}^{RR}\cdot}\ \bm{\eta}_{m}^{R}}_{Brownian}+\sum_{m\neq n}^{N}\,\sum_{l=1}^{\infty}\underbrace{\bm{\pi}_{nm}^{(R,\, l+1)}\odot{\bf V}_{m}^{(l+1)}+\bm{\Omega}_{n}^{a}}_{Active}
\end{alignat}
\end{subequations}\end{widetext}These equations are the main result of this section.
In the absence of activity, they reduce to the equation of Brownian dynamics with
hydrodynamic interactions \cite{ermak1978}. When the forces derive from positional
and angular potentials, the form chosen for the fluctuations ensures that the Gibbs
distribution of the positions and orientations is the stationary distribution. When
activity is included, the balance between fluctuation and dissipation is no longer
maintained and stationary states are no longer described by the Gibbs distribution.
As we show in the remainder of the paper, non-trivial stationary states are obtained
when the spheres are chained together into filaments.\foreignlanguage{american}{}
\begin{figure*}[t]
\includegraphics[width=1\textwidth]{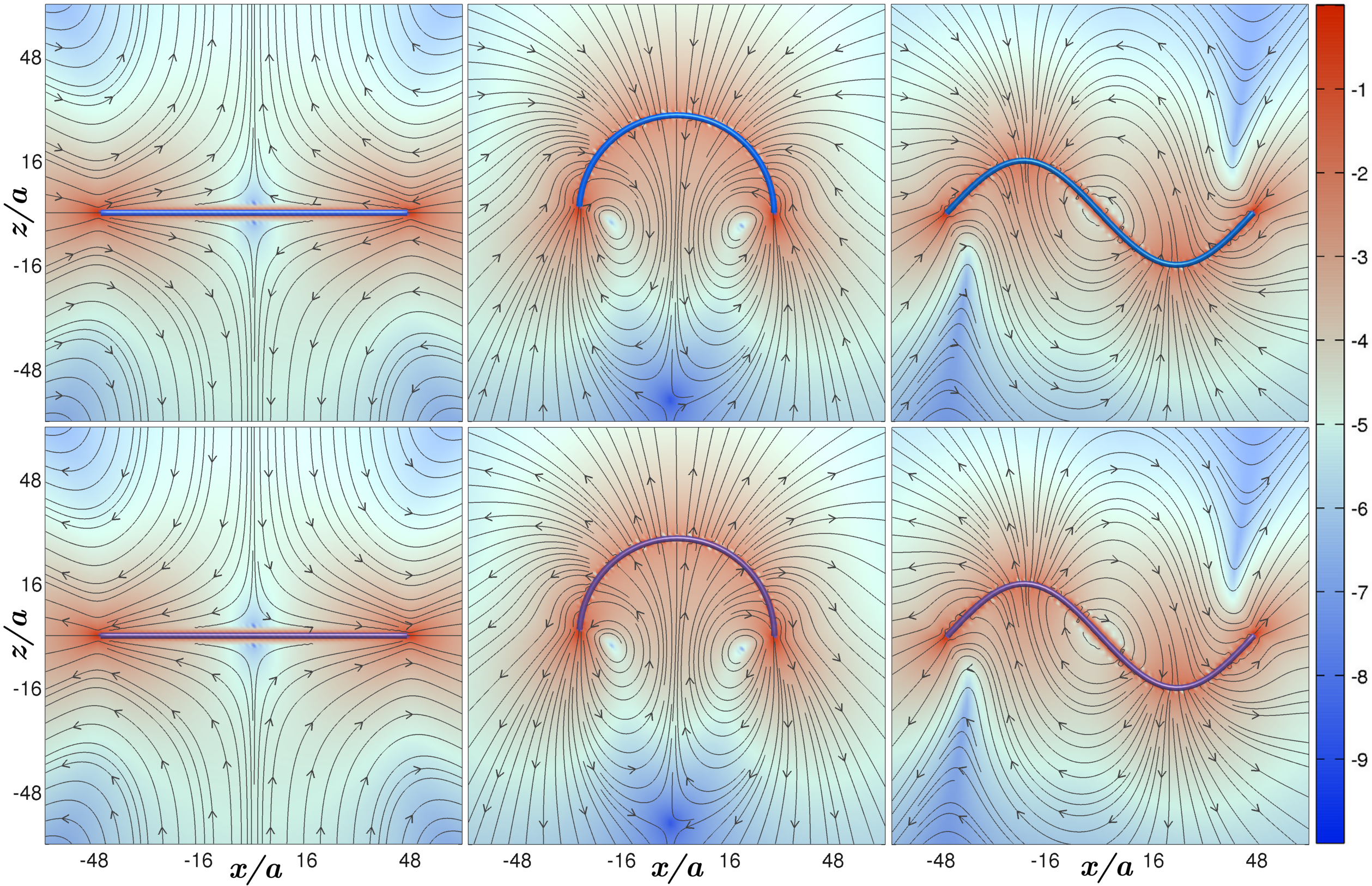}

\protect\caption{Streamlines of slow viscous flow around contractile (top row) and extensile (bottom
row) minimally active filaments. Spontaneous flow in the linear conformation tends
to compress contractile filaments and extend extensile filaments (first column).
Both symmetric and antisymmetric transverse perturbations are suppressed in contractile
filaments but enhanced in extensile filaments by the spontaneous flow (second and
third columns). The latter leads to a linear instability in extensile filaments,
when the elastic restoring force is no longer sufficient to counter the destabilizing
tendency of the spontaneous flow. }
 
\end{figure*}
\begin{figure*}[t]
\includegraphics[width=1\textwidth]{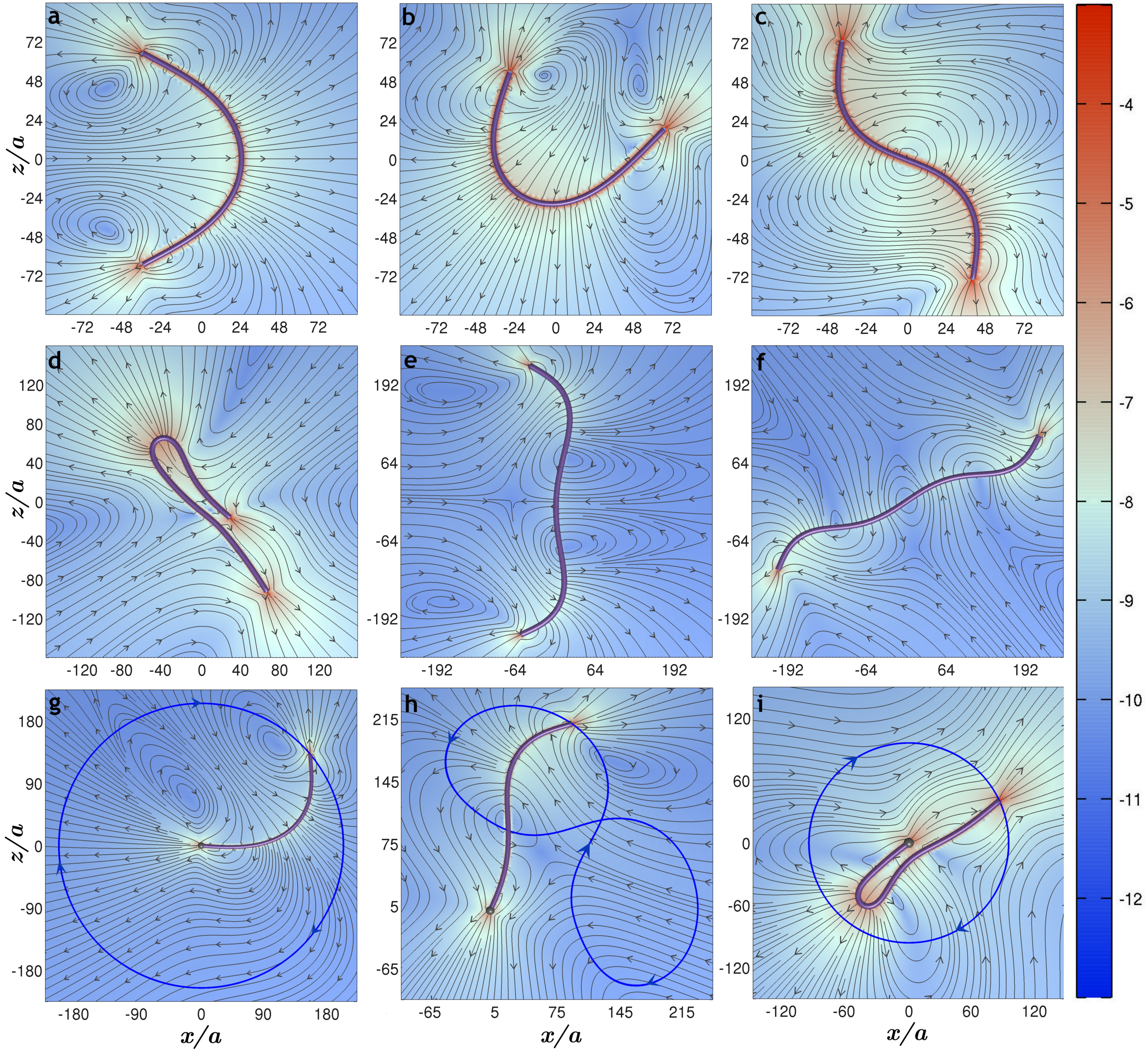}\protect\caption{Non-equilibrium stationary states of a free, (a) - (f), and tethered, (g) - (i),
minimally active filament. The figures show the filament conformation and the streamlines
of slow viscous flow produced by the activity, where the background is coloured by
the logarithm of the magnitude of the velocity. Free filament steady states are shown
for three lengths with increasing values of activity $\mathcal{A}_{s}$. For short
filaments ($N=48$) increasing activity produces the three non-equilibrium steady
states in (a) - (c), corresponding to the excitation of the first two elastic eigenmodes
and their linear combinations. As the filament length is increased ($N=80$ and $N=128$)
higher elastic eigenmodes appear with increasing amounts of activity in (d) - (f).
Tethered filament steady states are shown for a fixed length ($N=64$). Increasing
activity produces a rotating steady state in (g), which bifurcates into an oscillating
steady state in (h), with a return to a distinct rotating steady state in (i). (Movie)}
\end{figure*}

\section{Brownian microhydrodynamics of active filaments}

The equations of active Brownian hydrodynamics presented in the previous section
form the basis of our theory of active slender bodies. As mentioned before, we approximate
the slender body as a filament, and then discretize the filament as a set of connected
beads. The forces in the active Brownian hydrodynamic equations then are derivatives
of the non-linear spring potentials,
\begin{alignat*}{1}
{\bf F}_{n} & =-\bm{\nabla}_{n}U
\end{alignat*}
We assume the potential $U$ to be a sum of connectivity, elastic, and self-avoidance
potentials, 
\begin{alignat}{1}
U & =\sum_{m=1}^{N-1}U^{C}({\bf R}_{m},{\bf R}_{m+1})+\sum_{m=2}^{N-1}U^{E}({\bf R}_{m-1},{\bf R}_{m},{\bf R}_{m+1})\nonumber \\
 & +\sum_{m<n}U^{S}({\bf R}_{n},{\bf R}_{m})
\end{alignat}
The connectivity potential $U^{C}({\bf R},{\bf R}^{'})=k(r-b_{0})^{2}/2$, with elasticity
parameter $k$, penalizes departures of the distance, $r=|{\bf R}-{\bf R}^{'}|$,
of two consecutive particles from the equilibrium value of $b_{0}$. The elastic
potential for bending $U^{E}=\bar{\kappa}\left(1-cos\phi\right)$ , with rigidity
parameter $\bar{\kappa}$, penalizes departures of the angle $\phi$ between consecutive
bonds from their equilibrium value of zero. The rigidity parameter $\bar{\kappa}$
connects to bending rigidity $\kappa$ as $\bar{\kappa}=\kappa b_{0}$. The self-avoidance
potential $U^{S}$ restricts the overlap of particles and is taken here to be a Lennard-Jones
potential that vanishes smoothly at a distance $\sigma_{LJ}=2^{\frac{1}{6}}\sigma$.
The net force on the filament vanishes, as can be easily verified by summing the
force on each particle. 

An approximate method for computing the hydrodynamic interactions, used widely in
the dynamics of polymers, is to neglect their many-body character and, instead, assume
them to be pair-wise additive. This superposition approximation, first introduced
by Kirkwood and Riseman \cite{kirkwood1948intrinsic}, yields the well-known Rotne-Prager-Yamakawa
form for the translational mobility. In that approximation, the mobility matrices
of Eq. 1 are, \begin{subequations}

\begin{alignat}{1}
8\pi\eta\bm{\mu}_{nm}^{TT} & =\begin{cases}
\frac{4}{3}a^{-1}\bm{\delta} & \qquad\qquad\negthickspace m=n\\
\mathcal{F}^{0}\mathcal{F}^{0}\mathbf{G}(\mathbf{R}_{n},\mathbf{R}_{m}) & \qquad\qquad\negthickspace m\neq n
\end{cases}\\
8\pi\eta\bm{\mu}_{nm}^{TR} & =\begin{cases}
0 & \qquad\quad\negthickspace m=n\\
\frac{1}{2}\bm{\nabla}_{m}\times\mathbf{G}(\mathbf{R}_{n},\mathbf{R}_{m}) & \qquad\quad\negthickspace m\neq n
\end{cases}\\
8\pi\eta\bm{\mu}_{nm}^{RT} & =\begin{cases}
0 & \qquad\quad\negthickspace m=n\\
\frac{1}{2}\bm{\nabla}_{n}\times\mathbf{G}(\mathbf{R}_{n},\mathbf{R}_{m}) & \qquad\quad\negthickspace m\neq n
\end{cases}\\
8\pi\eta\bm{\mu}_{nm}^{RR} & =\begin{cases}
a^{-3}\bm{\delta} & \negthickspace m=n\\
\frac{1}{4}\bm{\nabla}_{n}\times\bm{\nabla}_{m}\times\mathbf{G}(\mathbf{R}_{n},\mathbf{R}_{m}) & \negthickspace m\neq n
\end{cases}
\end{alignat}
\end{subequations}

The diagonal parts of these matrices are one-body terms while the off-diagonal parts
represent the hydrodynamic interactions. The diagonal parts are the familiar Stokes
translational and rotational mobilities while the off-diagonal parts can be recognised
as the Rotne-Prager-Yamakawa tensors and their generalizations to rotational motion.
The Onsager symmetry of the mobility matrix is manifest in these expressions. In
the same approximation the propulsion matrices are \cite{singh2014many}\begin{subequations}
\begin{alignat}{1}
\negthickspace\negmedspace8\pi\eta\bm{\pi}_{nm}^{(T,\, l+1)} & =\begin{cases}
0 & m=n\\
c_{l}\mathcal{F}^{0}\mathcal{F}^{l}\mathbf{\bm{\nabla}}_{m}^{(l)}\mathbf{G}(\mathbf{R}_{n},\mathbf{R}_{m}) & m\neq n
\end{cases}\\
\negthickspace\negmedspace8\pi\eta\bm{\pi}_{nm}^{(R,\, l+1)} & =\begin{cases}
0 & \negthickspace\negmedspace m=n\\
\dfrac{c_{l}}{2}\bm{\nabla}_{n}\times\mathbf{\bm{\nabla}}_{m}^{(l)}\mathbf{G}(\mathbf{R}_{n},\mathbf{R}_{m}) & \negthickspace\negmedspace m\neq n
\end{cases}
\end{alignat}
\end{subequations}where, $\mathcal{F}^{l}$ is the operator
\begin{alignat*}{1}
\mathcal{F}^{l} & =\left(1+\frac{a^{2}}{4l+6}\bm{\nabla}^{2}\right).
\end{alignat*}
that accounts for the finite-size Faxen correction for the flow due to a sphere of
radius $a$ and $c_{l}$ is a numerical constant. The propulsion matrices, as defined
here, represent hydrodynamic interactions and, therefore, are purely off-diagonal.
The self-propulsion and self-rotation are wholly contained, respectively, in the
diagonal terms $\mathbf{V}_{n}^{a}$ and $\bm{\Omega}_{n}^{a}.$ 

For an active filament, the rotational motion of the spheres can be ignored, assuming
torsional potentials that prevent such rotations. Therefore, the orientation of the
spheres is no longer a dynamical degree of freedom to be determined from the angular
velocity, but is prescribed. It is natural to fix the orientation of $n$-th sphere,
$\mathbf{p}_{n}=\,\alpha\mathbf{t}_{n}+\beta\mathbf{n}_{n}+\gamma\mathbf{b}_{n}$,
in the local Frenet-Serret frame in terms of the of the tangent $\mathbf{t}_{n}$,
normal $\mathbf{n}_{n}$ and binormal $\mathbf{b}_{n}$, and the direction cosines
$\alpha$, $\beta$ and $\gamma$. Combining all of the above, we obtain the following
equation of motion for active filaments,\begin{widetext}

\begin{alignat}{1}
\dot{\mathbf{R}}_{n}=\, & \frac{\mathbf{F}_{n}}{6\pi\eta a}+\frac{1}{8\pi\eta}\sum_{m\neq n}^{N}\mathcal{F}^{0}\mathcal{F}^{0}\mathbf{G}(\mathbf{R}_{n},\mathbf{R}_{m})\cdot\mathbf{F}_{m}+\left(\frac{2k_{B}T}{6\pi\eta a}\right)^{\frac{1}{2}}\bm{\xi}_{n}+\sum_{m\neq n}^{N}\left(\frac{2k_{B}T\mathcal{F}^{0}\mathcal{F}^{0}\mathbf{G}(\mathbf{R}_{n},\mathbf{R}_{m})}{8\pi\eta}\right)^{\frac{1}{2}}\cdot\bm{\xi}_{m}\nonumber \\
+\, & \frac{1}{8\pi\eta}\sum_{m\neq n}^{N}\sum_{l=1}^{\infty}\underbrace{c_{l}\mathcal{F}^{0}\mathcal{F}^{l}\mathbf{\bm{\nabla}}_{m}^{(l)}\mathbf{G}(\mathbf{R}_{n},\mathbf{R}_{m})\odot\mathbf{V}_{m}^{(l+1)}+\mathbf{V}_{n}^{a}}_{Active}\label{eq:GeneralFilamentEquation1}
\end{alignat}
\end{widetext}These stochastic differential equations describe the Brownian dynamics
of an extensible, semi-flexible, self-avoiding active filament including hydrodynamic
interactions that arise from the exchange of momentum between the filament and its
local conservation in the bulk fluid. These equations are the natural extension of
Brownian hydrodynamics of passive filaments to the active case. In the next section,
we study the simplest version of these equations, where only the most dominant contribution
to active flow is retained and Brownian fluctuations are neglected.

\section{Dynamics Of Minimally Active Filaments}

In this section, we study in detail the simplest member of the family of models described
by the general equations of motion (Eq. \ref{eq:GeneralFilamentEquation1}) of the
previous section. In this minimal model, the spheres are non-motile, $\mathbf{V}_{n}^{a}=0$,
and all active velocity components other than the symmetric part of $\mathbf{V}_{n}^{(2)}$are
zero. Thus each individual sphere produces the flow show in Fig. 1b. The velocities
and tractions are, therefore, \begin{subequations}
\begin{alignat}{1}
\mathbf{v}(\mathbf{R}_{m}+\bm{\rho}_{m}) & =\dot{\mathbf{R}}_{m}+{\bf s}_{m}\cdot\bm{\rho}_{m}\\
4\pi a^{2}\mathbf{f}\,(\mathbf{R}_{m}+\bm{\rho}_{m}) & =-\bm{\nabla}_{m}U+3{\bf S}_{m}\cdot\bm{\rho}_{m}
\end{alignat}
\end{subequations}where $\mathbf{s}_{m}$ and $\mathbf{S}_{m}$ are, respectively,
the symmetric parts of the second-rank velocity and traction coefficients. The solution
of the boundary integral equation, in the diagonal approximation, relates the unknown
traction coefficient to the known value of the velocity coefficient \cite{singh2014many,ghose2014a}
\begin{gather}
{\bf S}_{m}=-\frac{3}{20\pi\eta a^{3}}{\bf s}_{m}
\end{gather}
To complete the model, it is necessary to specify the orientation $\mathbf{p}_{n}$
of the spheres, and hence the principal axis of the stresslet, in relation to the
filament conformation. Motivated by the experimental observation that molecular motors
walking on microtubules generate tangential stresses \cite{sanchez2011}, we parametrize
${\bf s}_{m}$ uniaxially, with its principal axis always parallel to the local tangent
${\bf t}_{m}$ of the filament,
\begin{gather}
{\bf s}_{m}=s_{0}(\mathbf{t}_{m}\mathbf{t}_{m}-\frac{1}{3}\bm{\delta}).
\end{gather}
The coefficient $s_{0}$ is positive for extensile stresses and negative for contractile
stresses. Additionally, we assume that the activity is so large that the Brownian
fluctuations make a negligible contribution to the dynamics. Active flow is balanced
entirely by the filament elasticity. This leads to deterministic equations of motion
for an active filament composed of non-motile beads,
\begin{alignat}{1}
\dot{\mathbf{R}}_{n}=-\frac{1}{6\pi\eta a}\bm{\nabla}_{n}U & -\frac{1}{8\pi\eta}\sum_{m\neq n}\mathcal{F}^{0}\mathcal{F}^{0}\mathbf{G}\cdot\bm{\nabla}_{m}U\nonumber \\
 & +\frac{7a^{3}}{6}\sum_{m\neq n}\underbrace{\mathcal{F}^{0}\mathcal{F}^{1}\bm{\nabla}\mathbf{G}\odot\mathbf{s}_{m}}_{Active}.
\end{alignat}
These equations, without finite-sized corrections to the hydrodynamic flow, were
first proposed in \cite{jayaraman2012} and subsequently used in \cite{laskar2013}
to study the dynamics of clamped active filaments. \foreignlanguage{american}{}
\begin{figure*}
\includegraphics[width=1\textwidth]{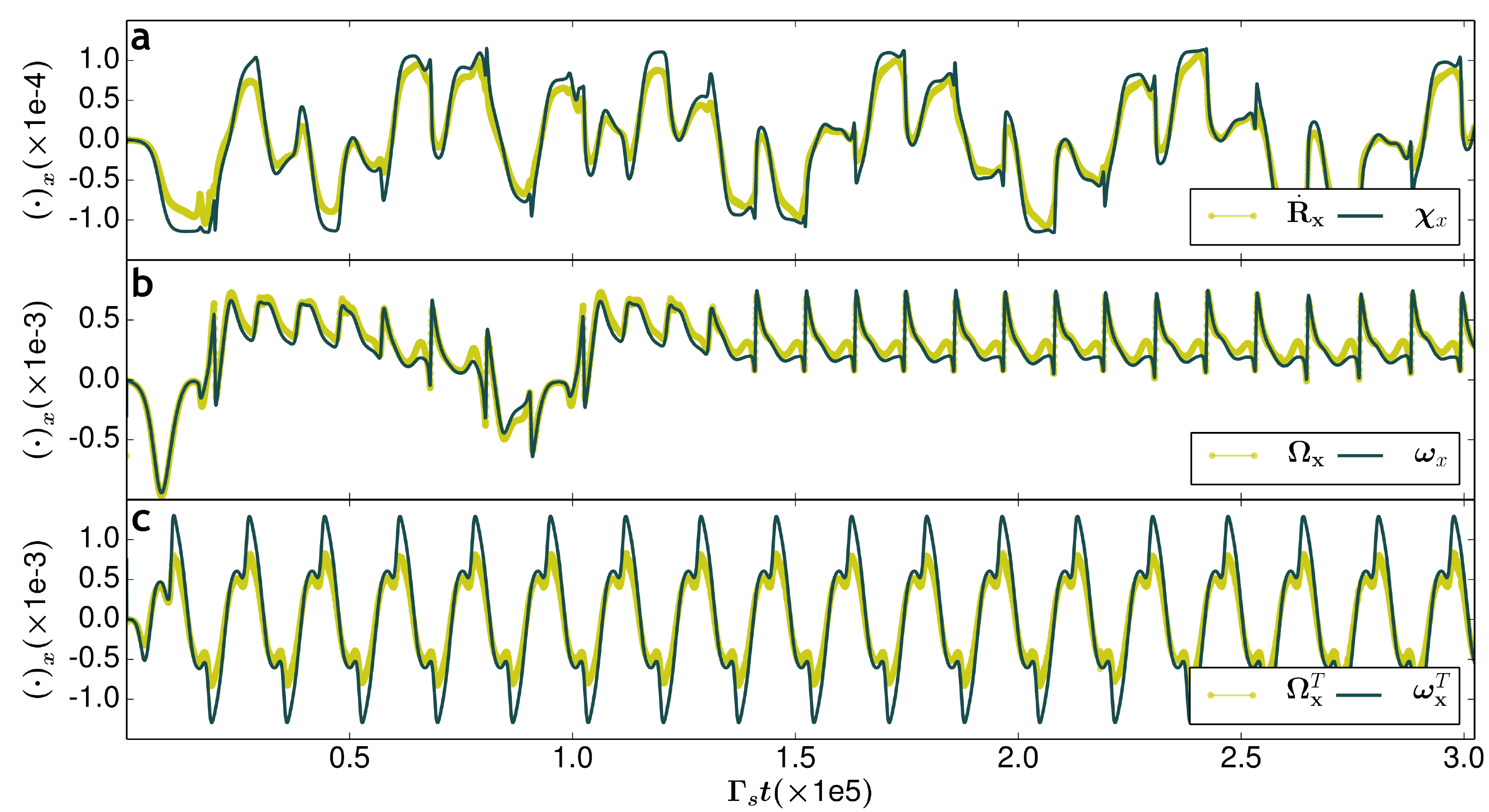}\label{fig:CurvatureMotionFree}

\protect\caption{Comparison of filament kinematics and curvature dynamics. In (a) the $x$-component
of the center of mass velocity is compared with the $x$-component of the scaled
average curvature vector (see Eq. 23) for a free filament. In (b) the $x$-component
of the mean angular velocity about the center of mas is compared with the scaled
average first moment of the curvature vector (see Eq. 24). In (c) the same comparison
is made for a tethered filament, with origin at the point of pivot. The comparison
shows that the filament motion is strongly correlated with dynamics of activity induced
curvature.}
\end{figure*}

The flow produced by the filament is sum of contributions from the potentials and
the activity, 

\begin{alignat}{1}
\negthickspace\mathbf{v}(\mathbf{r})=- & \frac{1}{8\pi\eta}\sum_{n=1}^{N}\mathcal{F}^{0}\mathbf{G}\cdot\bm{\nabla}_{n}U+\frac{7a^{3}}{6}\sum_{n=1}^{N}\underbrace{\mathcal{F}^{1}\bm{\nabla}\mathbf{G}\odot\mathbf{s}_{n}}_{Active}.
\end{alignat}
The resultant flow is shown for three conformations, for both extensile and contractile
filaments, in Fig 2. In the linear conformation, shown in the first column of Fig
2, the flow tends to compress contractile filaments and extend extensile filaments.
The stationary length of the filament, then, is somewhat shorter in the contractile
flow but somewhat longer in the extensile flow. In a symmetrically curved conformation,
shown in the second column of Fig, 2, the spontaneous flow tends to suppress curvature
in the contractile filaments but tends to enhance it for extensile filaments. The
suppression and enhancement is seen for antisymmetrically curved conformations in
the third row of Fig. 2. This shows that the interplay of flow and curvature is generally
stabilizing for contractile filaments while it is destabilizing for extensile filaments.
On dimensional grounds, a linear instability is expected when the filament length
$L>l_{\mathcal{A}}\sim\kappa/s_{0}$. In the remainder of the paper, we shall focus
only on extensile filaments and study the non-equilibrium stationary states that
appear as a consequence of the linear instability. 

In $d$ spatial dimensions, activity introduces a new rate $\Gamma_{s}=s_{0}/\eta L^{d}$
in addition to the rate of elastic relaxation $\Gamma_{\kappa}=\kappa/\eta L^{d+1}$
of the filament bending modes. The ratio, $\mathcal{A}_{s}=Ls_{0}/\mathcal{\kappa}$,
of these two time scales provides a dimensionless measure of activity. The activity
number $\mathcal{A}_{s}$ also measures the departure from equilibrium and the amount
of energy is injected into the fluid by the filament. We vary both the filament length
and the activity number in studying the dynamics of the filament in $d=3$ dimensions. 

We integrate the equations of filament equations of motion numerically using a variable
coefficient method. We chose the following parameters for the model : spring constant
$k=1$, bondlength $b_{0}=4a$, rigidity parameter $\bar{\kappa}=0.4$, stresslet
strength $s_{0}=0.0-0.5$ and number of spheres $N=32-128$. We obtain the mobility
and propulsion matrices using the PyStokes library \cite{singh2014pystokes}. We
simulate the system for several hundred active relaxation time scales with an initial
condition that is linear with small random, transverse perturbations. We study two
cases, the first in which the filament is free at both ends and second in which it
is free at one end and tethered at the other end.

Our results are summarised in Fig. 3, which shows the non-equilibrium stationary
states for both free filaments in panels (a) - (f) and for tethered filaments in
panels (g) - (i). A linear instability appears at $\mathcal{A}_{s}\sim12$ in free
filaments which leads to spontaneous symmetric curvature and an emergent autonomous
motility, shown in panel (a). The conformation corresponds, roughly, to the first
elastic eigenmode of the passive filament. With increasing amounts of activity, higher
elastic eigenmodes appear through a series of bifurcations, shown in panels (b) -
(f). Whenever the conformation is asymmetrical about the center, the filament acquires
a rotational component of motion. The higher elastic eigenmodes appear for smaller
values of activity in longer filament, as is seen by comparing panel (c) with panels
(e) and (f). Although the system is three-dimensional, filament motion is planarly
stable, in a plane that is determined by the initial condition. This rich dynamics
(movie I) and non-equilibrium steady sates we have found is remarkably similar to
an experiment on an isolated axoneme \cite{bessen1980}. 

Tethering the filament at one end, restricts translation and thus the energy transduction
from the activity is fed into rotational and oscillatory states (Movie II). Beyond
the threshold activity of $\mathcal{A}_{s}\sim5$, a tethered filament rotates around
the pivot, panel (g), in a plane that is chosen by the initial condition. Further
energy injection leads to flagella like beating in a plane, shown in panel (h). The
highest value of activity studied is shown in panel (i), where a conformation similar
to panel (d) appears, but now constrained by the pivot, is forced to rotate while
maintaining conformation. With increase in filament length and activity, we expect
higher elastic eigenmodes to appear, and either rotate or oscillate under the constraint
of the tether. 

The dynamics of the center of mass follows from summing the equation of motion over
all spheres. The contribution from internal spring forces vanishes, and on approximating
the active flow by its contribution from the nearest neighbours, an approximate equation
is obtained that relates the center of mass motion to the filament curvature and
the activity,
\begin{alignat}{1}
\mathbf{V}_{CM} & \simeq-\frac{s_{0}}{4\pi\eta b_{0}}\langle\varkappa\hat{{\bf n}}\rangle=\mathbf{\bm{\chi}}
\end{alignat}
In Fig. 4a, we compare the numerically computed values of the center of mass velocity
with the curvature vector $\mathbf{\bm{\chi}}$ defined above. There is a surprisingly
good agreement between the two, indicating that principal effect of the non-local
active flow can be expressed locally as a tendency to promote curvature. A similar
relation holds for the angular velocity about the center of mass,
\begin{alignat}{1}
\bm{\Omega}_{CM}\simeq & -\frac{s_{0}}{8\pi\eta b_{0}}\langle\varkappa\hat{{\bf n}}\times({\bf R}-{\bf R}_{CM})\rangle=\bm{\omega}
\end{alignat}
 and the previous comparison is repeated for both free and tethered filaments in
Fig 4b and Fig 4c. For tethered filaments the pivot point, and not the center of
mass, is used to compute cross products, and the activity-dependent prefactor is
two times smaller. This suggests that effective, local equations of motion may be
accurate for describing some aspects of the dynamics of active filaments. 
\begin{figure*}
\selectlanguage{american}%
\subfigure[Free filament with and without HI]{\foreignlanguage{english}{\includegraphics[width=0.99\textwidth]{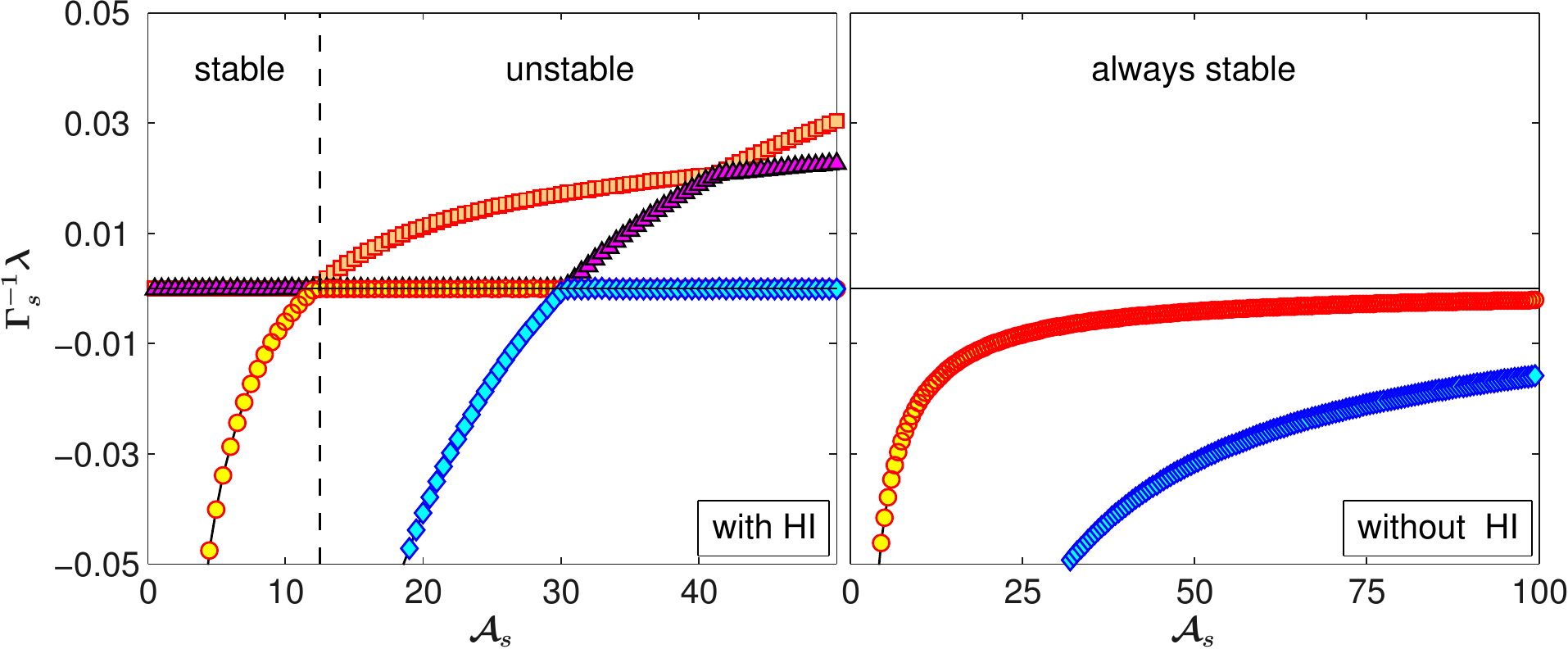}}}\\ \subfigure[Tethered Filament with and without HI]{\foreignlanguage{english}{\includegraphics[width=0.99\textwidth]{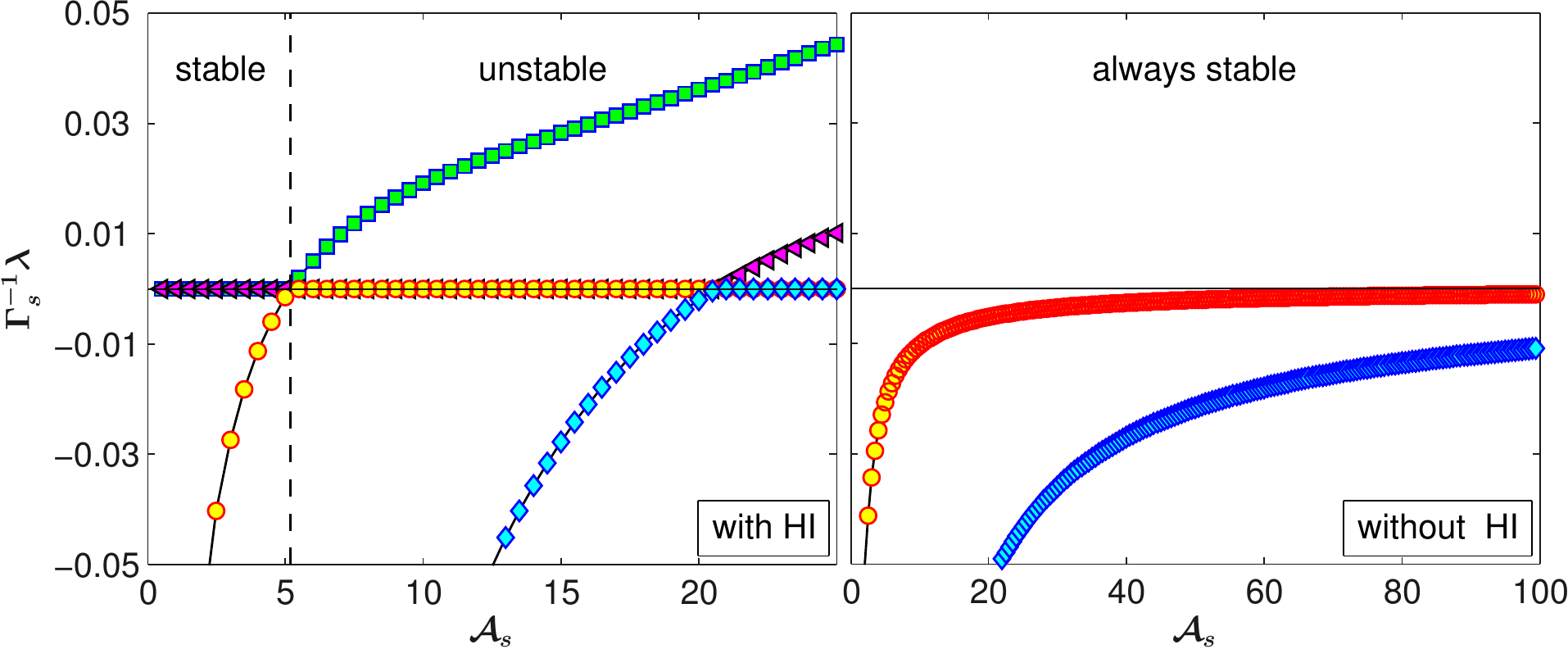}}}\foreignlanguage{english}{\protect\caption{Variation of the largest non-zero eigenvalues of the stability matrix, $\mathbf{J}$,
plotted against activity number, $\mathcal{A}_{s}$, for a free filament in (a) and
a tethered filament in (b). In each case, the eigenvalues are computed including
full hydrodynamic interactions (HI) (left panels) and neglecting all non-local hydrodynamic
contributions (right panels). The eigenvalues remain negative, for all values of
$\mathcal{A}_{s}$ in a large range, when hydrodynamic interactions are neglected.
Hydrodynamic interactions, therefore, are essential for the instability of the linear
conformation and the bifurcation to dynamical steady states, signalled by the positive
eigenvalues in each of the left panels. }
}\selectlanguage{english}%
\end{figure*}

\section{Linear Stability Analysis}

To better understand the linear instability which is expected from the flows shown
in Fig 2, we perform a stability analysis of the equations of motion, about the linear
conformation. Taking the equations of motion to represent a dynamical system, $\dot{\mathbf{R}}_{n}=f(\mathbf{R}_{1},\mathbf{R}_{2},\cdots,\mathbf{R}_{n})$,
we compute the Jacobian $\mathbf{J}=-\bm{\nabla}_{n}f\Big|_{\mathbf{R}_{n}^{0}}$
at the stationary state with linear conformation $\mathbf{R}_{n}^{0}$. The linearised
dynamics, then, is
\begin{gather}
\delta\dot{\mathbf{R}}_{n}=\mathbf{J}\cdot\delta\mathbf{R}_{n}
\end{gather}
 We numerically compute the eigenvalues of this stability matrix as a function of
activity $\mathcal{A}_{s}$ for both free and tethered filaments. To evaluate the
importance of non-local hydrodynamic interactions, we also compare the eigenvalues
for the dynamics in which all non-local (that is $m\neq n$) terms are deleted. The
results are shown in Fig. 5(a) and Fig. 5(b) for free and tethered filaments, respectively. 

We see that the largest eigenvalue becomes positive at $\mathcal{A}_{s}\sim12$ for
free filaments and $\mathcal{A}_{s}\sim5$ for tethered filaments. The bifurcation
is thus a simple instability and not of the Hopf type reported in the previous study
of a clamped minimally active filament \cite{laskar2013}. The first eigenmode instability
flows to the symmetrically curved conformation shown in Fig 3 (a). Instabilities
of the higher eigenmodes leads, the first of which is visible in Fig. 5a at $\mathcal{A}_{s}\sim40$,
produces the more complicated states shown in panels (b) - (g) of Fig 5. The presence
of the tether increases the threshold value of the activity at which the instability
occurs to $\mathcal{A}_{s}\sim5$, but the sequence of instabilities remains identical.
Remarkably, there is no instability, in the same range of activity, when hydrodynamic
interactions are ignored, as shown in the right panels of Fig 5. Thus, non-local
active hydrodynamic flow is essential to produce the instabilities and the non-equilibrium
stationary states reported above. 

Subsequent bifurcations with increasing values of activity are expected to have a
more complicated character, as the stationary states are typically limit cycles.
The numerical study of limit cycle instabilities is considerably more involved than
that of time-independent stationary states. We shall explore this aspect of the dynamics
of active filaments in a future study.

\section{Discussion and Conclusion}

Previous work on bead-spring models of active filaments have focused on three distinct
mechanisms of activity. In the earliest work of Jayaraman et al \cite{jayaraman2012},
activity arises from the hydrodynamic flow of the active spheres. In that work, the
equations of motion for filament dynamics in three dimensions contained contributions
from the leading order hydrodyamic flow due to stresslets and degenerate quadrupoles.
A detailed study and all results were given for non-motile active spheres, thus ignoring
the velocity quadrupoles. In subsequent work, Chellakot et al \cite{chelakkot2014flagellar}
studied a chain of motile active spheres, subject to Brownian motion, in two dimensions
but ignored the all non-local hydrodynamic effects, both passive and active. In related
work, Jiang and Hou \cite{jiang2014hydrodynamic} studied a chain of passive spheres,
subject to forces of non-equilibrium origin, directed along the filament tangent.
In their model, both passive hydrodynamic flow and hydrodynamically correlated Brownian
motion is included in three dimensions, but active flow is absent. Remarkably, in
spite of these differences between the models, they yield a broadly similar phenomenology
: linear instabilities, spontaneous motion, and oscillatory states. 

To understand why this might be, it is best to situate all the previous models within
the equations of motion presented here. The model studied in detail by Jayaraman
et al \cite{jayaraman2012} is obtained when self-propulsion velocities, $\mathbf{V}_{n}^{a},$
are set to zero, only the dipolar contribution to active flow is retained, and finite-sized
corrections to hydrodynamic flow as well as Brownian motion are neglected. The model
of Chellakot et al \cite{chelakkot2014flagellar} is obtained when the self-propulsion
velocity is directed along the axis $\mathbf{p}_{n},$ $\mathbf{V}_{n}^{a}=v_{s}\mathbf{p}_{n}$,
and this axis is itself determined from the balance of a restoring and Brownian torques.
This requires the angular velocity to be retained as a dynamical variable and all
off-diagional contributions to mobility and propulsion matrices to be ignored. Finally,
the model of Jiang and Hou \cite{jiang2014hydrodynamic} is obtained by ignoring
all active components of flow, $\mathbf{v}^{a}=0$, but representing the force on
the spheres as ${\bf F}_{n}=-\bm{\nabla}_{n}U+\alpha\mathbf{t}_{n},$ as sum of contributions
from the potentials and an unspecified non-equilibrium source. The common feature
of all these models is that they produce motion in the direction of the curvature.
This arises from the non-local hydrodynamic flow in the model of Jayaraman et al,
and from the local contributions due to self-propulsion and non-equilibrium activity
in the models of Chellakot et al and Jiang et al respectively. The present work shows
that a phenomenology beyond curvature instabilities remains to be explored. In particular,
torsional instabilities, possible with self-rotating active spheres that are unhindered
by torsional potentials, are likely to yield further surprises in the dynamics of
active filaments. 

The equations of motion presented provide the foundation for studying non-equilibrium
statistical mechanics of active filaments. The coupled Langevin equation for the
positions can be recast as Fokker-Planck equations, whose stationary solutions in
the absence of activity are given by the Gibbs distribution. Activity, in the forms
envisaged in this work, introduces an additional drift in the Fokker-Planck equation,
destroying the balance between fluctuation and dissipation. This will lead to non-Gibbsian
distributions in the stationary state, and, likely change well-known equilibrium
properties like statics of the coil-globule transition \cite{kaiser2015does} and
the distribution of loop closure times \cite{Snigdha2014RingClosure}. We urge that
some of these problems be addressed both experimentally and through theory and simulations. 

As a final remark, we draw attention to the similarity between the instabilities
reported here and the convective instability by active stress studied in the pioneering
work of Finlayson and Scriven \cite{finlayson1969}. 
\begin{acknowledgments}
RA wishes to thank D. Frenkel, M. E. Cates, P. Chaikin, G. Date, A. Donev, S. Ghose,
A. J. C. Ladd, I. Pagonabarraga, R. Singh, R. Singh, D. J. Pine, M. J. Shelley, H.
A. Stone and P. B. Sunil Kumar for many useful discussions while this work was being
completed. RA acknowledges the Hamied Trust for supporting a visit to Cambridge University
and the IUSSTF for supporting visits to Princeton University and NYU and warmly thanks
D. Frenkel, H. A. Stone and D. J. Pine and for hosting at the respective universities.
The authors gratefully acknowledge the Department of Atomic Energy, Government of
India for supporting their research. 
\end{acknowledgments}

\appendix
\begin{widetext}

\section{Integrals and Matrix elements }

The expressions for the boundary integrals $\mathbf{G}^{(l)}$ and $\mathbf{K}^{(l)}$
and the matrix elements $\mathbf{G}_{nm}^{(l,l')}$ and $\mathbf{K}_{nm}^{(l,l')}$
are given as \cite{singh2014many},

\begin{subequations}
\begin{alignat}{1}
\mathbf{G}^{(l+1)}(\mathbf{r},\mathbf{R}_{m}) & =\frac{2l+1}{4\pi a^{2}}\int\mathbf{G}(\mathbf{r},\mathbf{R}_{m}+\bm{\rho}_{m})\mathbf{Y}^{(l)}(\hat{\bm{\rho}}_{m})dS_{m}=a^{l}\bm{\Delta}^{(l)}\mathcal{F}^{l}\mathbf{\bm{\nabla}}_{m}^{(l)}\mathbf{G}(\mathbf{r},\mathbf{R}_{m})\\
\mathbf{K}^{(l+1)}(\mathbf{r},\mathbf{R}_{m}) & =\frac{1}{l!(2l-1)!!}\int\mathbf{K}(\mathbf{r},\mathbf{R}_{m}+\bm{\rho}_{m})\cdot\mathbf{n}\mathbf{Y}^{(l)}(\hat{\bm{\rho}}_{m})dS_{m}=\frac{4\pi a\: a^{l}\bm{\Delta}^{(l)}}{(l-1)!(2l+1)!!}\mathcal{F}^{l}\mathbf{\bm{\nabla}}_{m}^{(l-1)}\mathbf{K}(\mathbf{r},\mathbf{R}_{m})
\end{alignat}
\end{subequations}\begin{subequations}

\begin{alignat}{1}
\mathbf{G}_{nm}^{(l+1,l'+1)}(\mathbf{R}_{n},\mathbf{R}_{m}) & =\begin{cases}
{\displaystyle \delta_{ll'}\frac{2l+1}{2\pi a}\int\mathbf{Y}^{(l)}(\hat{\bm{\rho}})\left(\bm{\delta}-\hat{\bm{\rho}}\hat{\bm{\rho}}\right)\mathbf{Y}^{(l)}(\hat{\bm{\rho}})\, d\Omega;} & \qquad\quad\;\:{\displaystyle m=n;}\\
\, & \,\\
{\displaystyle a^{l+l'}\mathcal{F}_{n}^{l}\mathcal{F}_{m}^{l'}\bm{\nabla}_{n}^{(l)}\bm{\nabla}_{m}^{(l')}\mathbf{G}(\mathbf{R}_{n},\mathbf{R}_{m});} & {\displaystyle \qquad\quad\;\:{\displaystyle m\neq n};}
\end{cases}\\
\mathbf{K}_{nm}^{(l+1,l'+1)}(\mathbf{R}_{n},\mathbf{R}_{m}) & =\begin{cases}
-{\displaystyle \delta_{ll'}4\pi\bm{\delta}\bm{\Delta}^{(l)};} & {\displaystyle m=n;}\\
\, & \,\\
{\displaystyle \frac{4\pi a^{(l+l'+1)}}{(l'-1)!(2l'+1)!!}\mathcal{F}_{n}^{l}\mathcal{F}_{m}^{l'}\bm{\nabla}_{n}^{(l)}\bm{\nabla}_{m}^{(l'-1)}\mathbf{K}(\mathbf{R}_{n},\mathbf{R}_{m});} & {\displaystyle m\neq n;}
\end{cases}
\end{alignat}
\end{subequations}

\section{Constraining Force on the tethered bead}

Tethering the filament generates a conformation dependent constraint force at the
tethered point. We take care of hydrodynamics to satisfy the boundary condition appropriately,
such that net velocity of the respective end vanishes. Then we back-calculate the
constrained force self-consistently on the course of simulation. 

\begin{flalign*}
\dot{\mathbf{R}}_{1}= & -\frac{1}{6\pi\eta a}\left(\bm{\nabla}_{1}U+\mathbf{F}_{c}\right)-\frac{1}{8\pi\eta}\sum_{m=2}^{N}\mathcal{F}^{0}\mathcal{F}^{0}\mathbf{G}(\mathbf{R}_{1},\mathbf{R}_{m})\cdot\bm{\nabla}_{m}U+\frac{7a^{3}}{6}\sum_{m=2}^{N}\mathcal{F}^{0}\mathcal{F}^{1}\bm{\nabla}\mathbf{G}(\mathbf{R}_{1},\mathbf{R}_{m})\odot\mathbf{s}_{m}\\
\mathbf{F}_{c} & =-\bm{\nabla}_{1}U-\frac{3}{4\eta}\sum_{m=2}^{N}\mathcal{F}^{0}\mathcal{F}^{0}\mathbf{G}(\mathbf{R}_{1},\mathbf{R}_{m})\cdot\bm{\nabla}_{m}U+7\pi\eta a^{4}\sum_{m=2}^{N}\mathcal{F}^{0}\mathcal{F}^{1}\bm{\nabla}\mathbf{G}(\mathbf{R}_{1},\mathbf{R}_{m})\odot\mathbf{s}_{m}.
\end{flalign*}

\end{widetext}

\bibliographystyle{unsrt}

\end{document}